%% file: paper.tex
\documentclass[twocolumn]{aastex631}
\usepackage{import}

\newcommand{\rmunit}{\,rad\,m$^{-2}$}
\newcommand{\psrchive}{{\tt PSRCHIVE}}
\newcommand{\ter}{Terzan~5}

\shorttitle{Pulse Profiles and Polarization of Terzan~5 Pulsars}
\shortauthors{Martsen et al.}

\graphicspath{figures/}

\begin{document}

\title{Radio Pulse Profiles and Polarization of the \ter\ Pulsars}

\author[0000-0001-8313-0895]{Ashley R.~Martsen}
\affiliation{Rochester Institute of Technology, 1 Lomb Memorial Dr., Rochester, NY 14623, USA}

\author[0000-0001-5799-9714]{Scott M.~Ransom}
\affiliation{National Radio Astronomy Observatory, 520 Edgemont Road, Charlottesville, VA 22903, USA}

% Authors beyond this point are in alphabetical order by surname

\author[0000-0002-2185-1790]{Megan E.~DeCesar}
\affiliation{George Mason University, Fairfax, VA 22030, resident at the U.S. Naval Research Laboratory, Washington, D.C. 20375, USA}

\author[0000-0003-1307-9435]{Paulo C.~C.~Freire}
\affiliation{Max-Planck-Institut für Radioastronomie, Auf dem Hügel 69, D-53121 Bonn, Germany}

\author[0000-0003-2317-1446]{Jason W.~T.~Hessels}
\affiliation{ASTRON, the Netherlands Institute for Radio Astronomy, Oude Hoogeveensedijk 4, 7991 PD Dwingeloo, The Netherlands}
\affiliation{Anton Pannekoek Institute for Astronomy, University of Amsterdam, Science Park 904, 1098 XH, Amsterdam, The Netherlands}

\author[0000-0002-9017-3567]{Anna Y. Q.~Ho}
\affiliation{Department of Astronomy, Cornell University, Ithaca, NY 14853, USA}

\author[0000-0001-5229-7430]{Ryan S.~Lynch}
\affiliation{Green Bank Observatory, PO Box 2, Green Bank, WV 24494, USA}

\author[0000-0001-9784-8670]{Ingrid H.~Stairs}
\affiliation{Dept.~of Physics and Astronomy, UBC, 6224 Agricultural Road, Vancouver, BC V6T 1Z1 Canada}

\author[0000-0001-5538-0395]{Yuankun Wang}
\affiliation{Dept.~of Astronomy, Box 351580, University of Washington, Seattle, WA 98195, USA}

\begin{abstract}

\ter\ is a rich globular cluster within the galactic bulge containing 39 known millisecond pulsars, the largest known population of any globular cluster. These faint
pulsars do not have enough signal-to-noise (S/N) to measure reliable flux density or polarization information from individual observations in general. We combined over 5.2\,days of archival data, at 1500\,MHz and 2000\,MHz, taken with the Green Bank Telescope over the past 12\,years. We created high S/N profiles for 32 of the pulsars and determined precise rotation measures (RMs) for 28. We used the RMs, pulsar positions, and dispersion measures (DMs), to map the projected parallel component of the Galactic magnetic field toward the cluster. The $\langle B_{||}\rangle$ shows a rough gradient of $\sim$6\,nG/arcsec ($\sim$160\,nG/parsec), or fractionally, a change of $\sim$20$\%$ in the right ascension direction across the cluster, implying Galactic magnetic field variability at sub-parsec scales. We also measured average flux densities $S_\nu$ for the pulsars, ranging from $\sim$10\,$\mu$Jy to $\sim$2\,mJy, and an average spectral index $\alpha = -1.35$, where $S_\nu \propto \nu^{\alpha}$. This spectral index is flatter than most known pulsars, likely a selection effect due to the high frequencies used in pulsar searches to mitigate dispersion and scattering. We used flux densities from each observation to constrain the scintillation properties towards the cluster, finding strong refractive modulation on months timescales. The inferred pulsar luminosity function is roughly power-law, with slope $(d\log N)/(d\log L) = -1$ at the high-luminosity end. At the low-luminosity end, there are incompleteness effects implying that \ter\ contains many more pulsars.

\end{abstract}


\section{Introduction} \label{sec:intro}

Globular clusters are excellent environments for the formation of millisecond pulsars (MSPs). The high density of stars cause stellar interactions that can then produce binaries with neutron stars, and those neutron stars can be recycled to become MSPs \citep[e.g.][]{Ransom_2008}. \ter\ is one of the most massive, dense, and metal-rich globular clusters in the galaxy \citep{Ter5_structure, Ter5_structure2}, located at distance  $D=6.62\pm$0.15\,kpc from Earth towards the Galactic Center, and embedded in the Galactic Bulge \citep{Gaia_2021}.

\ter\ currently has 39 known pulsars, all of which are recycled, and the vast majority of which are true millisecond pulsars (i.e.~$P$ $<$10\,ms) \citep{PulsarDisc1990,PulsarDisc1998, ransom2001,PulsarDisc2005,Hessels2006,PulsarDiscaj,PulsarDiscam,2021Meerkat}.  The pulsars have been used to study the physical characteristics of the cluster \citep{prager17} and have also been the subject of deep X-ray studies \citep[e.g.][]{Bogdanov}, since the cluster also contains many interesting high-energy emitting binaries due to its very large stellar interaction rate.  Most of \ter's pulsars are extremely faint in the radio band with flux density of tens of $\mu$Jy, and some were found only due to refractive scintillation which made them detectable in certain observations (see \S\ref{ssec:scint}). Even multi-hour observations with the 100-m Green Bank Telescope (GBT) often result in very low signal-to-noise detections of many of the pulsars, and sometimes even non-detections for the faintest. In this paper we combine the detections from hundreds of hours of GBT observations to make much higher signal-to-noise flux- and polarization-calibrated pulse profiles for 32 of the pulsars in the cluster.

Their extreme faintness
makes polarization studies very difficult, since the percentage of linearly polarized signal over the total intensity for MSPs is typically 10$-$40\%, with occasional pulsars having more and others having less polarization \citep[e.g.~][]{Gentile2018,wahl2021nanograv}. By carefully summing together many individually calibrated observations, we can measure the average pulse shapes much more precisely (which improves pulsar timing), and also increase the signal-to-noise of the polarized components of the emission.  For most of the last 12 years, we have been recording data with full polarization information (Stokes parameters I, Q, U and V) % PF: defined it here, as you use the letters I, Q, U and V after this without definition
with the GBT for \ter, and in this paper use those data to determine the polarized pulse profiles including total linear $L = \sqrt{Q^2 + U^2}$ component, circular $V$ component, and the position angle using the IAU/IEEE convention of the linear polarization, $\Phi = 1/2\arctan\left(U/Q\right)$.

The position angles of the linear polarization are rotated as they travel through the interstellar medium due to Faraday rotation, which is observable thanks to its frequency dependence. This allows us to determine the rotation measure (RM) of the pulsars, if they have enough linear polarization and our observing band is wide enough. Much like the dispersion measure (DM) is the line-of-sight election column density, RM is the same but also weighted by the line-of-sight parallel component of the magnetic field strength, and encodes the strength of the Faraday rotation \citep[e.g.][]{wahl2021nanograv}. Understanding the RM can help inform our understanding of the Galactic magnetic fields between \ter's pulsars and us, since the averaged parallel component of the intervening magnetic field is  \begin{equation}
    \label{B_par}
    \langle B_{||} \rangle = 1.23\,\mathrm{\mu G} \left( \frac{\mathrm{RM}}{\mathrm{rad}\,\mathrm{m}^{-2}} \right) \left( \frac{\mathrm{DM}}{\mathrm{pc}\,\mathrm{cm}^{-3}} \right)^{-1}
    \label{eqn:bpar}
\end{equation}
\citep{pulsar_handbook}.

Since \ter\ is a distant globular cluster, the angular separations between the pulsar pairs are on the order of arcseconds (and all the pulsars are effectively at the same distance), allowing extremely rare small scale mapping of the Galactic magnetic fields \citep[e.g.][]{Tuc47}. In this paper we provide the summed total intensity and polarized profiles of the \ter\ pulsars, measurements of both their average flux densities and their flux densities over time, spectral indices, and RMs, and determine a sparse map of $\langle B_{||} \rangle$ towards the cluster.

\section{Observations and Data Analysis} \label{sec:observations}

We used archival data from GBT observations from 2010 October to 2021 May, with cadences typically roughly monthly during the early years and shifting to roughly quarterly in the later years. Two different receivers and observing bands were used: the L-band receiver, centered at 1500\,MHz and the S-band receiver, centered at 2000\,MHz, each of which had $\sim$650\,MHz of usable bandwidth out of 800\,MHz provided by the observing systems. Two different pulsar instruments were used to record the data: GUPPI \citep{GUPPI} and VEGAS \citep{VEGAS}. GUPPI was used from 2010 until it was replaced by VEGAS in 2019 December.  Both pulsar ``backends'' sampled two orthogonal linear polarizations with 8-bit samplers, and coherently dedispersed the data at a dispersion measure (DM) of 238\,cm$^{-3}$\,pc, which is near the average value of the \ter\ pulsars.  The data were written into search-mode PSRFITS\footnote{\url{https://www.atnf.csiro.au/research/pulsar/psrfits_definition/Psrfits.html}} files with 512 frequency channels at a time resolution of 10.24\,$\mu$s in the ``coherence'' format (i.e.~with both self- and cross-products of the two polarizations).

While the GUPPI and VEGAS data have the same center frequencies and total instrumental bandwidths, GUPPI records its data in ``upper sideband'' while VEGAS uses ``lower sideband'', meaning that the frequency directions of the channels across the observing bands are stored in opposite manners.  In addition, while they both record data in ``coherence'' format, the polarization conventions in the data headers are different, with GUPPI using the IAU/IEEE standard and VEGAS using the PSR/IEEE standard \citep{2010psrchive}, causing differences in the recovered Stokes parameters when processed with {\tt PSRCHIVE}.  Finally, the first channel recorded by each instrument is contaminated due to the way they compute spectra via discrete Fourier transforms.  That channel is discarded for both instruments, resulting in 511 overlapping channels between the instruments.  These differences make combining folded pulsar data from the GUPPI and VEGAS instruments impossible with current {\tt PSRCHIVE} tools.

We folded the coherently dedispersed search-mode data using {\tt fold\_psrfits} from the {\tt psrfits\_utils} package\footnote{\url{https://github.com/scottransom/psrfits_utils}}, with timing ephemerides derived with {\tt TEMPO} and updated every few years via standard techniques.  Integrations were recorded for most pulsars every five minutes (the exceptions being one minute integrations for Ter5A and three minute integrations for Ter5I), resulting in four-dimensional data cubes (time, polarization, observing frequency, and pulse phase) stored in PSRFITS format. We calibrated the data for flux and the two leading polarization terms (differential gains and delays of the two polarizations) using one-minute duration pulse calibrator observations taken before each multi-hour (typically 4$-$8 hours in duration) monitoring observation.  The pulsed calibrator observations were themselves calibrated using semi-regular observations of the quasars B1442$+$101 or 3C190.  Post-calibration, we removed radio frequency interference primarily using automated algorithms in {\tt PSRCHIVE}'s {\tt paz} routine, specifically via the median smoothed difference algorithm and subinterval ``lawn mowing'' algorithms, with the {\tt -r} and {\tt -L} flags, respectively.  We examined each folded observation of each pulsar manually and, if necessary, applied additional manual RFI excision.  Before we performed the detailed summation procedure described in \S\ref{ssec:summation}, we installed the most recent timing ephemerides into each of the PSRFITS archives and multiplied all flux densities by a constant factor of 20 since the FPGA code that GUPPI and VEGAS uses for coherently dedispersed search-mode data internally divides the data by a factor of 20 to prevent overflows during FPGA signal processing.

In total, there were 31 L-band observations (28 with GUPPI and 3 with VEGAS), and 39 S-band observations (30 with GUPPI and 9 with VEGAS).  However, many of the observations after 2014 March taken with GUPPI were usable only for total intensity work and not polarization analysis due a randomly occurring timing instability between the two polarizations on one of the GUPPI sampler boards \citep{wahl2021nanograv}.  Manual examination of the Stokes $V$ data for several of the brighter pulsars with significant circular polarization (e.g.~Ter5A and N) identified which of the GUPPI observations were usable for polarization work since the timing anomaly causes ripples in $V$ across the observing band due to uncorrected rotation measure from linear polarization being mixed into Stokes $V$ from Stokes $U$. For the GUPPI data, 20 of 28 L-band observations and 9 of 30 S-band observations were usable for polarization work.  All of the observations were used for the total intensity profiles.

Of the 39 pulsars in \ter, we created summed profiles for all but seven. We excluded Ter5A since it exhibits orbital phase wander and erratic eclipses, making alignment and summation extremely difficult.  However, since it is a fairly bright pulsar, we used strong individual observations, with non-eclipsed data aligned and co-added to produce high quality total intensity and polarization profiles. Since the eclipsing material can cause de-polarization \citep[e.g.~][]{Ter5A_RM}, we used only the very centers of the non-eclipsed portions of several orbits from one L-band day (2013 January 1) and one S-band day (2013 April 17) for our analyses. We excluded Ter5P and Ter5ad since, similar to Ter5A, they are ``redbacks'' and show strong and erratic eclipses, orbital variability, and occasionally months of non-detections \citep[e.g.~][]{Hessels2006}, making accurate profile alignment and summation difficult.  We also excluded the recently discovered pulsars Ter5aj through Ter5an 
\citep{PulsarDiscaj,PulsarDiscam,2021Meerkat} since they are all quite faint, and more importantly, most of the earlier high-resolution and full Stokes data had been permanently deleted before they were discovered due to its large data volume (720\,GB\,hr$^{-1}$).

\subsection{Summation Method} \label{ssec:summation}

We co-added the L-band and S-band observations using \psrchive's {\tt psradd} routine. For the total intensity profiles we included all observations, and for the polarization profiles we excluded data from corrupted dates. Once the observations had been calibrated and cleaned as described in \S\ref{sec:observations}, we installed the best known timing ephemeris and assumed-constant DM into each observation for a given pulsar, ensuring that we used the barycentric frequency correction for each observation (i.e.~with {\tt Dispersion::barycentric\_correction = 1} in the {\tt PSRCHIVE} configuration file), this allowed us to integrate each observation in time (i.e.~with {\tt pam -T } from \psrchive). We combined each set of observations using the same backend and receiver using the \psrchive\ routine {\tt psradd} along with the {\tt -P} option, which aligns each observation based on the location of the integrated pulse profile, which in general produced a better alignment than using the ephemeris. This is likely due to an imperfect ephemeris and the complicated environment that the pulsars are found in. {\tt psradd} combines the data using a weighting based on the radiometer equation, such that the weights are proportional to $1/\sigma^2$, where $\sigma \propto (T_\mathrm{obs}\,\mathrm{BW})^{-1/2}$,  $T_\mathrm{obs}$ is the non-RFI-zapped duration of the observation, and $\mathrm{BW}$ is the non-RFI-zapped radio bandwidth.

\subsection{Instrument Combination} \label{ssec:combination}

The previously described summation process resulted in eight sets of combined data files for each pulsar,  those with good polarization data or total intensity for both GUPPI and VEGAS observations at both L-band and S-band. We used the L-band GUPPI total intensity sums to find new DM values for each pulsar using {\psrchive}'s {\tt{pdmp}}, and then manually adjusted the DM by eye, if needed, so that the leading edges of each summed and high signal-to-noise profile were aligned across the band when displayed via {\tt pav -GTpd}, which shows pulse phase versus observing frequency. These new ``best DMs'', (shown in Table~\ref{tab1}), are specified with five significant figures since that precision lets us align all of the pulse profiles across L-band to 1{\%} or better, for even the fastest pulsars in \ter. We installed these DMs into the original datafiles and performed the previous summation one final time.  This last iteration improved the signal-to-noise of the final sums for many of the faintest pulsars, and especially for several which had less accurate DMs previously due to low signal-to-noise.

Given the differences in the GUPPI and VEGAS data described in \S\ref{sec:observations}, we were unable to use {\tt PSRCHIVE} to directly combine the observations taken with those two instruments. We therefore wrote a Python routine using the {\tt PSRCHIVE} python interface to change the VEGAS data into a format identical to that of the more common GUPPI data.  As mentioned in \S\ref{sec:observations}, the main differences between the GUPPI and VEGAS data are that they are stored in opposite sideband and with two differing polarization conventions, IAU/IEEE and PSR/IEEE respectively  \citep{2010psrchive,wahl2021nanograv}.  The Python routine band-flipped the usable 511 channels of the VEGAS data, flipped the signs of Stokes $Q$, $U$, and $V$ (which accounts for the different circular polarization and position angle conventions between the PSR/IEEE and IAU/IEEE conventions), and then combined those 511 channels and polarizations with the GUPPI data using the same radiometer equation based weighting as described in \S\ref{ssec:summation}.

\subsection{Rotation Measure Determination} \label{ssec:RMs}

We used the combined L-band observations with good polarization to measure RM values for each pulsar with two independent routines, {\tt rmfit} from \psrchive\ and {\tt fit\_RM.py} from {\tt RMNest} \citep[][]{Bannister19,Lower20}\footnote{\url{https://github.com/mlower/rmnest}}.  With {\tt rmfit}, which determines the RM by maximizing the amount of linear polarization as a first step, we used the {\tt -r} and {\tt -W } options, which divides the band into two equally-weighted intervals (based on wavelength-squared) and refines the determined RM and error estimates based on the position angles determined in the two separate bands.  {\tt fit\_RM.py} uses a Bayesian technique based on nested-sampling to determine the best RM and confidence intervals.  The results from the two fits are given in Table~\ref{tab1}, and are consistent for most of the pulsars. For pulsars with multiple linearly-polarized components with significant phase separation (e.g.~Ter5O, V, and ae), we also used {\tt RMNest} on the individual components to investigate the consistency of the RM measurements.We note that the error estimates seem to be systematically larger for {\tt RMNest}, which also seems to be able to determine an RM for pulsars with smaller fractional amounts of linear polarization than {\tt rmfit}.  There is an additional component of the measured RM from the Earth's ionosphere during each of the observations.  We estimated the size of those contributions with the \texttt{ionFR}\footnote{\url{https://github.com/csobey/ionFR}} code, and they had a mean of about $+$2.7\,\rmunit\ and a standard deviation of 1.5\,\rmunit\ \citep{ionFRpaper}.  Since we did not correct the individual measured RMs for ionospheric contributions, there is likely a systematic bias of positive $2-3$\,\rmunit\ for each of the measurements.  However, those biases should be the same for each pulsar, and so relative comparisons of the RMs should be unaffected. The summed and RM-corrected pulse profiles at L-band are shown in Fig.~\ref{fig:Profiles}.

\subsection{Average Flux Density Determination} \label{ssec:fluxes}

Using the instrument-combined total-intensity summed profiles at L-band and S-band, we determined the average flux densities and spectral indices for each pulsar.  We first made noiseless templates from the summed L-band and S-band profiles using Gaussian fitting with the {\tt pygaussfit.py} routine from {\tt PRESTO} \citep[][]{presto}.  Those noiseless templates were used both to update the timing as described in \S\ref{ssec:timing}, and to determine flux densities of 100-MHz-wide subbands using \psrchive's {\tt psrflux} routine.  Since the L-band and S-band data overlap by 300\,MHz, we fit the resulting 16 flux density measurements ($S_\nu$) as a function of observing frequency, $\nu$, to a power-law model ($S_\nu\propto\nu^\alpha$) referenced to 1400\,MHz, with a pulsar-specific constant offset also fitted between L-band and S-band. The results of the fits are in Table~\ref{tab1}. We note that the power-law model was not an exceptionally good fit for about a quarter of the pulsars (see the reduced-$\chi^2$ fit values), and so the errors on the measured spectral indices $\alpha$ are likely underestimated by tens of percent for those pulsars. The average of the offset between observing bands was 1.17 (i.e.~decreasing the fluxes at S-band by that factor) with a standard deviation across the pulsars of $\sim$0.20, implying that on average, either the L-band data had underestimated flux densities or that the S-band data had overestimated flux densities by that amount.

\subsection{Flux Density versus Time} \label{ssec:SvT}

Since the GBT observations were a mix of L-band and S-band data, in order to study the variability of the pulsars as a function of time, we measured flux densities for each day, and then used the spectral indices from \S\ref{ssec:fluxes} to convert that flux density to 1750\,MHz for each observation. That frequency is the center of the overlapping 300\,MHz portions of the two observing bands.  In order to help eliminate differences in calibration over time and between the two observing bands, we effectively normalized the measured flux densities for a day using the ensemble of flux densities of all the pulsars on that day compared to their individual weighted average values over time.  Specifically, if we know the weighted average flux density of each pulsar, we can determine an average scale factor for each day that corrects the individual pulsar flux densities to their long-term weighted average.  Averaging the pulsar scale factors allows the individual pulsars to vary, yet doesn't bias measurements of the bright pulsars like a constant total flux density constraint would.

\section{Results} \label{sec:results}

\begin{figure*}[h]
        \includegraphics[width=7in]{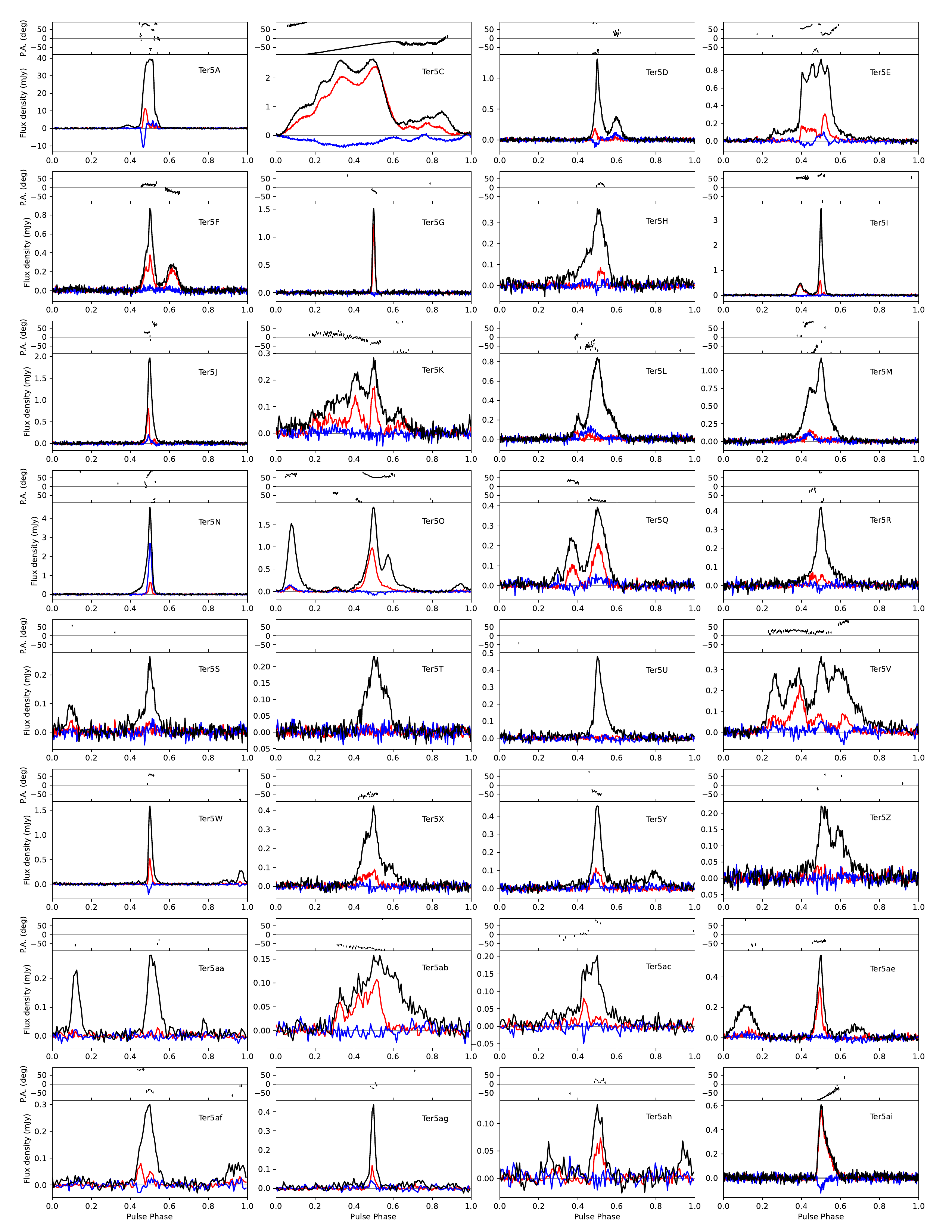}
        \caption{Summed profiles for 32 of the L-band (covering roughly 1150$-$1850\,MHz) profiles for the \ter\ pulsars. The remaining pulsars are not shown due to very low signal-to-noise, or complications that kept them from being summed. The profile from Ter5A is from a single observation (see text).
        The black lines are total intensity, red lines are the linear polarization $L$ with the mean $L$ off-pulse subtracted, and the blue is the circular polarization $V$ in the IAU/IEEE convention. The position angle is shown above each profile if it is $3\sigma$ significant, resulting in some anomalous ``detections'' due to noise even in off-pulse portions of the profiles. The primary y-axis scales (in flux density, mJy), are different for each pulsar.}
        \label{fig:Profiles}
\end{figure*}

For each pulsar we ended with four different summed pulse profiles:  the combined L-band and S-band data either with full Stokes information or with just total intensity (i.e.~using all of the GUPPI and VEGAS data).  In total, the summed polarization profiles included roughly 71.3\,hours of observations at S-band and 137.4\,hours of L-band. The total intensity profiles contained 124.9\,hours of S-band data and 157.8\,hours of L-band. The final L-band polarization profiles for the pulsars we analyzed are shown in Fig.~\ref{fig:Profiles}. 

\begin{figure}[h]
        \includegraphics[width=3.4in]{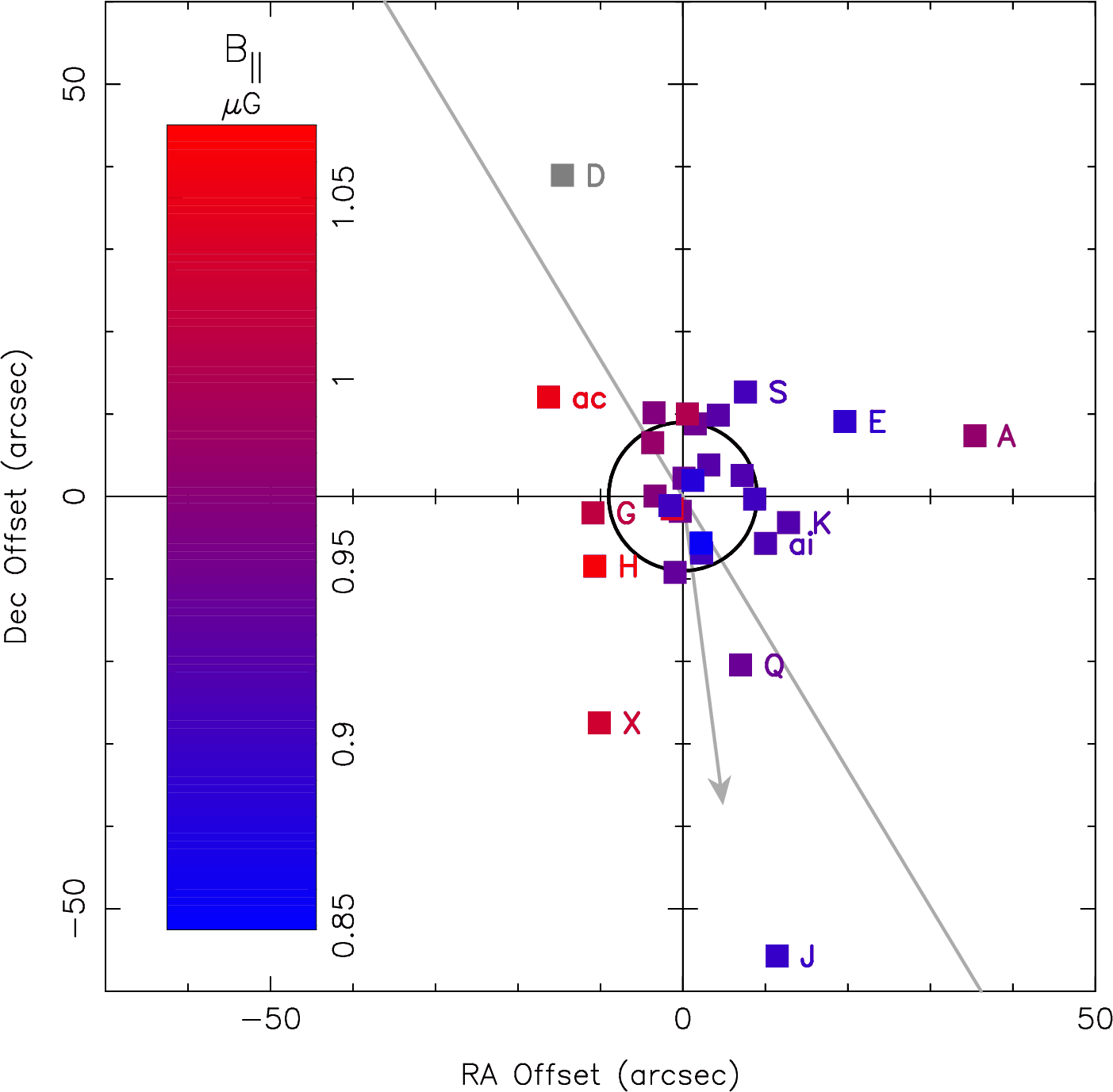}
        \caption{A map of the projected average parallel component of the Galactic magnetic field strength towards Terzan~5, centered on the optical position of the cluster, with the core radius shown as a black circle \citep{Ter5_structure}. The grey arrow points to the Galactic center, which is 4.19\,$^{\circ}$ in that direction, and the grey line through the center of the cluster is parallel to the Galactic plane. The color bar indicates $\langle B_{||} \rangle$ in $\mu$G towards each pulsar, from the lowest value to the second highest. These values were calculated using the RMs as determined by {\tt rmfit } (see Table~\ref{tab1}). We omitted Ter5aa since the RM values determined by the two different methods are inconsistent, and greyed-out Ter5D since it has a significantly higher RM (and therefore $\langle B_{||}\rangle$) than the rest of the pulsars, $\sim$1.18\,$\mu$G. There is an obvious gradient in $\langle B_{||} \rangle$ roughly in the Right Ascension (RA) direction.}
        \label{fig:B_par}
\end{figure}

The DM values we used to sum the data for each pulsar, the two RM measurements, and the measured flux densities at 1400\,MHz and 2000\,MHz as well as the spectral index, $\alpha$, are shown in Table~\ref{tab1}. Together, we used the RM and DM measurements to calculate the magnetic field parallel to the line of sight $\langle B_{||} \rangle$ (see Equation~\ref{eqn:bpar}), which is mapped in Fig.~\ref{fig:B_par}.

Using the known 6.62$\pm$0.15\,kpc distance ($D$) to \ter\ \citep{Gaia_2021}, we converted the L-band flux densities ($S_\mathrm{1400}$) into pseudo-luminosities ($L_\mathrm{1400} = S_\mathrm{1400} D^2$, assuming isotropic beaming) and plotted them as a luminosity function in Fig.~\ref{fig:Lumin}.

The 1750\,MHz flux densities and their errors as a function of time, as determined in \S\ref{ssec:SvT}, for Ter5C, O, Z, and ae are shown in Fig.~\ref{fig:Flux_density}.  We also used the 1750\,MHz time series to see if scintillation from the intervening interstellar medium might cause correlations in the pulsar flux densities.  We computed correlation coefficients between each pair of pulsars using the Spearman rank-order technique, and the average coefficient was -0.03$\pm$0.16.  Furthermore, there were no cases of strong correlations, even between the three pairs of pulsars separated by $<$1$\arcsec$ (Ter5I, Z, and ae).  For example, the correlation coefficient for the closest pair, Ter5ae and Ter5Z, separated by only $\sim$0.47$\arcsec$ is -0.16, which is not statistically significant. The pulsar flux densities do vary significantly, though, by factors of up to $\sim$2. The average modulation index at 1750\,MHz, $m = \sigma_S / \bar S$, where $\sigma_S$ is the standard deviation of the flux density measurements and $\bar S$ is their average, is 0.25$\pm$0.04.

\section{Discussion} \label{sec:Discussion}

\subsection{Profiles}
\label{ssec:Profiles}

As can be seen from the profiles in Fig.~\ref{fig:Profiles}, there is a wide variety of pulse shapes and polarization fractions. Due to the fact that all pulsars were observed in the same data and calibrated with the same method, the differences are due to differences between the pulsars themselves and not due to observational effects. To list some interesting profiles; Ter5N is highly circularly polarized, and is the only pulsar in \ter\ with more circular polarization than linear polarization. Ter5C has an incredibly wide pulse and nearly no off-pulse region, and is strongly polarized over the entire pulse. Ter5ai is nearly $100\%$ linearly polarized. We can use this knowledge to confirm that the low levels of polarization in some pulsars is not due to incorrect calibrations, since this effect would be the same in all the pulsars. This means since we know the polarization calibrations should be correct, we can say that Ter5U has almost no polarization of any form. Ter5G is a very narrow pulse with very strong linear polarization and no circular polarization, whereas most other pulsars tend to have some amount of circular polarization or else their linear polarization is weak. Ter5V has a wide pulse as well, and interestingly the position angle is nearly flat across most of the profile. The last interesting pulsar to highlight is Ter5O, which has a strong and complicated pulse shape and polarization.

\subsection{Rotation Measures and Parallel Magnetic Field}
\label{ssec:BField}

While the majority of pulsars did not previously have measured RMs, the RM for Ter5A was measured by both \citet{BilousThesisRM} and \citet{Ter5A_RM}. We determined an RM of 191.2(3)\rmunit\ and 193(2)\rmunit\ for {\tt rmfit} and {\tt RMNest}, respectively, which is consistent with the value determined by \citet{BilousThesisRM}, 190(10)\rmunit. \citet{Ter5A_RM} measured 174.9(4.2)\rmunit\ at L-band with the Parkes telescope, which is significantly different from our measurement.  However, due to the eclipsing nature of Ter5A, the varying ionized wind from the companion star causes changes in the polarization properties of the pulsar signal as a function of both time and orbital phase, including significant changes in the instantaneous observed RM.

Using RM values from Galactic field pulsars or AGN to measure the parallel magnetic field has been attempted a number of times, \citep[e.g.~][]{Han_pulsar_old,b_field_AGN, Han_pulsar_recent,  b_field_old, b_field}. This works very well on a large scale, but due to the separations between the sources, this method usually probes scales between degrees or 10s of arcminutes in $\langle B_{||} \rangle$. Using globular cluster pulsars, as opposed to AGN or Galactic pulsars, the separations between pulsars are on the scale of tens of arcseconds. Therefore globular cluster pulsars can trace the changes in $\langle B_{||} \rangle$ on the same scales of arcseconds (which translates to pc or fractions of pc in linear scales). Another benefit of using globular cluster pulsars, as opposed to Galactic pulsars, is that the distance to all the cluster pulsars are effectively the same, allowing for direct comparisons between RMs and $\langle B_{||} \rangle$ between the different sight lines.

A study completed by \citet{Tuc47} used the pulsars in 47~Tucanae to constrain the Galactic magnetic field. However, a major difference between the 47~Tuc study and this one on \ter, are the locations of the clusters within or near the Milky Way. 47~Tuc is in the halo of the Galaxy, well off of the Galactic plane, and due to the minimal ionized interstellar medium (ISM) between the cluster and us, the differences in $\langle B_{||} \rangle$ for the pulsars could be attributed to the magnetic field within 47~Tuc. For \ter, because it is embedded in the Galactic disk and located toward the Galactic center, the intervening ISM dominates the DMs and RMs of the pulsars \citep[see][]{prager17}, making it impossible to study the cluster magnetic field itself. The variations that we observe in $\langle B_{||}\rangle$ are likely due to small scale changes in the Galactic magnetic field and/or variations (i.e.~turbulence) in the ionised ISM between \ter\ and us.

As shown in Fig.~\ref{fig:B_par}, we measure significant variations in $\langle B_{||}\rangle$ across the cluster, including a gradient  of roughly $\sim$6\,nG/arcsec or $\sim$160\,nG/parsec along the RA direction, which corresponds to a fractional change of about 20\% across the cluster. The RM values used for Fig.~\ref{fig:B_par} were calculated using {\tt{rmfit}}, which in general, agreed with those determined using {\tt RMNest} (see Table \ref{tab1}). There were two pulsars, however, Ter5C and D, which did not agree within 2$-\sigma$ between the two codes. Also, Ter5aa is only within 2$-\sigma$ due to a very large uncertainty in the {\tt RMNest} value.

The systematic errors in the RM values for the pulsars are expected to be $\sim$1\,\rmunit, caused by varying but averaged atmospheric contributions. The measured RM values found for Ter5C are consistent within the expected systematic error.  The disagreement is likely due to the very small statistical uncertainties, combined with calibration and fitting complications due to the highly complex pulse profile of Ter5C.
Ter5D has the highest measured RM of the \ter\ pulsars, and it is also furthest from the center of the cluster. Ter5aa has very little linear polarization, which likely affects the estimates of the RMs and the errors.

Along with the disagreement between values from {\tt{rmfit}} and {\tt{RMNest}}, we were unable to measure the RMs of several pulsars using one or both methods. {\tt{rmfit}} was unable to determine RMs for Ter5S, Ter5U, or Ter5Z, while neither method was able to measure the RM for Ter5T. Each of these four pulsars have extremely small (or negligible) amounts of linear polarization.

Our RM analysis of the phase-separated polarized pulse components of Ter5O, V, and ae using {\tt RMNest} was intriguing, but inconclusive.  The independent RMs from the pulse and interpulse of Ter5O were consistent within their errors and also with the pulse-averaged RM given in Table \ref{tab1}.  However, the different components for Ter5V and ae provided statistically discrepant RMs by several sigma, with absolute values similar to the RM values in Table \ref{tab1}. It is hard to conclude much from these preliminary results given the low signal-to-noise of the summed profiles. Independent RM measurements of different pulse components from much brighter MSPs seems warranted.

\begin{figure}[h]
        \includegraphics[width=3.7in]{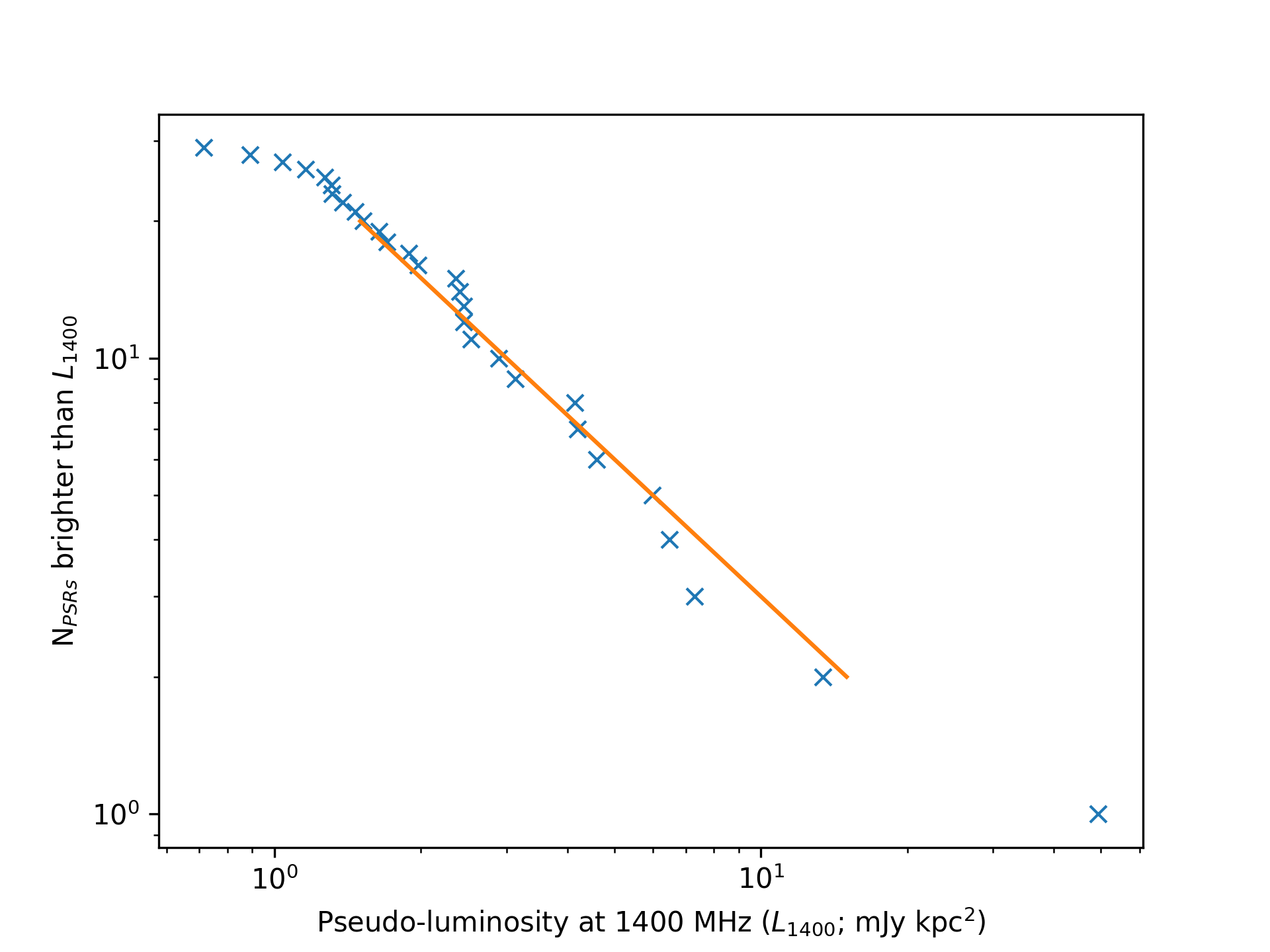}
        \caption{Pseudo-luminosity ($L_{1400}$) vs cumulative number of pulsars with pseudo-luminosity greater than that luminosity. A line with slope of $-$1 (not a fit) has been plotted in orange to compare to the trend of the data.  A turnover at low luminosity is evident below $\sim$1\,mJy\,kpc$^2$, implying incompleteness in our searches, and that additional pulsars likely exist in the cluster.}
        \label{fig:Lumin}
\end{figure}

\subsection{Pulsar Timing}
\label{ssec:timing}

Due to the faintness and complex pulse profiles of the \ter\ MSPs, a substantial benefit of the new integrated pulse profiles is much more accurate pulse profile templates to be used for the timing of the pulsars.  Better templates provide more accurate and more precise arrival times (TOAs) which directly lead to more accurate and more precise timing solutions.  We have tested the potential timing improvements for 29 of the \ter\ pulsars where fitting a series of Gaussians to the integrated profiles makes a better template than simply smoothing those same profiles (the brighter pulsars A, C, and O benefit from the latter).  For the 29 pulsars, 28 show improvements in TOA precision, with an average improvement of the median TOA precision of $\sim$34\%, and with many of the pulsars with weak and/or complex profiles (D, E, K, L, U, V, W, X, Y, Z, and ag) having precision improvements of $\gtrsim$50\%  (i.e.~factors of 2$-$3.5 smaller error bars).  Meanwhile, basically all of those pulsars also saw similar improvements in the accuracy of the TOAs as determined by lower RMS-values for the timing residuals or improved reduced-$\chi^2$ of the timing fits.  The only pulsars without substantial improvements (i.e.~$\lesssim$10\%) were those with relatively simple profiles which we had, by luck, been effectively modeling previously with only one or two Gaussians (G, H, J, T, and ai).  The full results of the timing will be presented elsewhere.

\subsection{Flux Densities and Luminosities}
\label{ssec:Flux_Disc}

By measuring the average flux densities of the pulsars over $\sim$1400\,MHz of bandwidth, and also over scores of observations which average out scintillation effects, we were able to make good measurements of the spectral indices of the pulsars, assuming that their flux densities varied as a power law (see \S\ref{ssec:fluxes}). The average spectral index of the pulsars is $-$1.35 (with a standard deviation of 0.53), which is significantly flatter than has been measured for other pulsars \citep[e.g.][]{bilous16,jankowski18, Aggarwal22}.  In particular, \citet{ng12.5narrow} measured the spectral indices of 49 MSPs (although 2 are measured twice, independently) and those have a mean spectral index of about $-$1.58.  For \ter\, the flatter than typical spectral index is highly likely biased due to the fact that most of the pulsars in the cluster were found at S-band \citep[e.g.][]{PulsarDisc2005,Hessels2006}.  Due to the rapidly increasing effects of scattering and pulse-broadening from the intervening ISM at lower radio frequencies, it is likely that multiple steep-spectrum pulsars have been missed in searches of the cluster.

We can also use the measured flux densities and the known distance to the cluster to compute the pseudo-luminosity function for the \ter\ pulsars (see Fig.~\ref{fig:Lumin}).  The data at higher luminosities closely follows a log-log slope of $-$1, and a turnover is apparent at lower luminosities implying incompleteness in our searches.  As many MSPs exist with 1400\,MHz luminosities at a level of 0.1\,mJy\,kpc$^2$ or below \citep[e.g.][]{psrcat}\footnote{\url{http://www.atnf.csiro.au/research/pulsar/psrcat}}, it seems likely that there are dozens, or perhaps even $>$100 additional pulsars still to be found in \ter, in agreement with {\em Fermi} LAT data \citep{Gamma_number} and earlier radio studies \citep{hessels07,gclumin2011,gclumin2013,tl13}. 

\subsection{Scintillation}
\label{ssec:scint}
\begin{figure}[h]
        \includegraphics[width=3.7in]{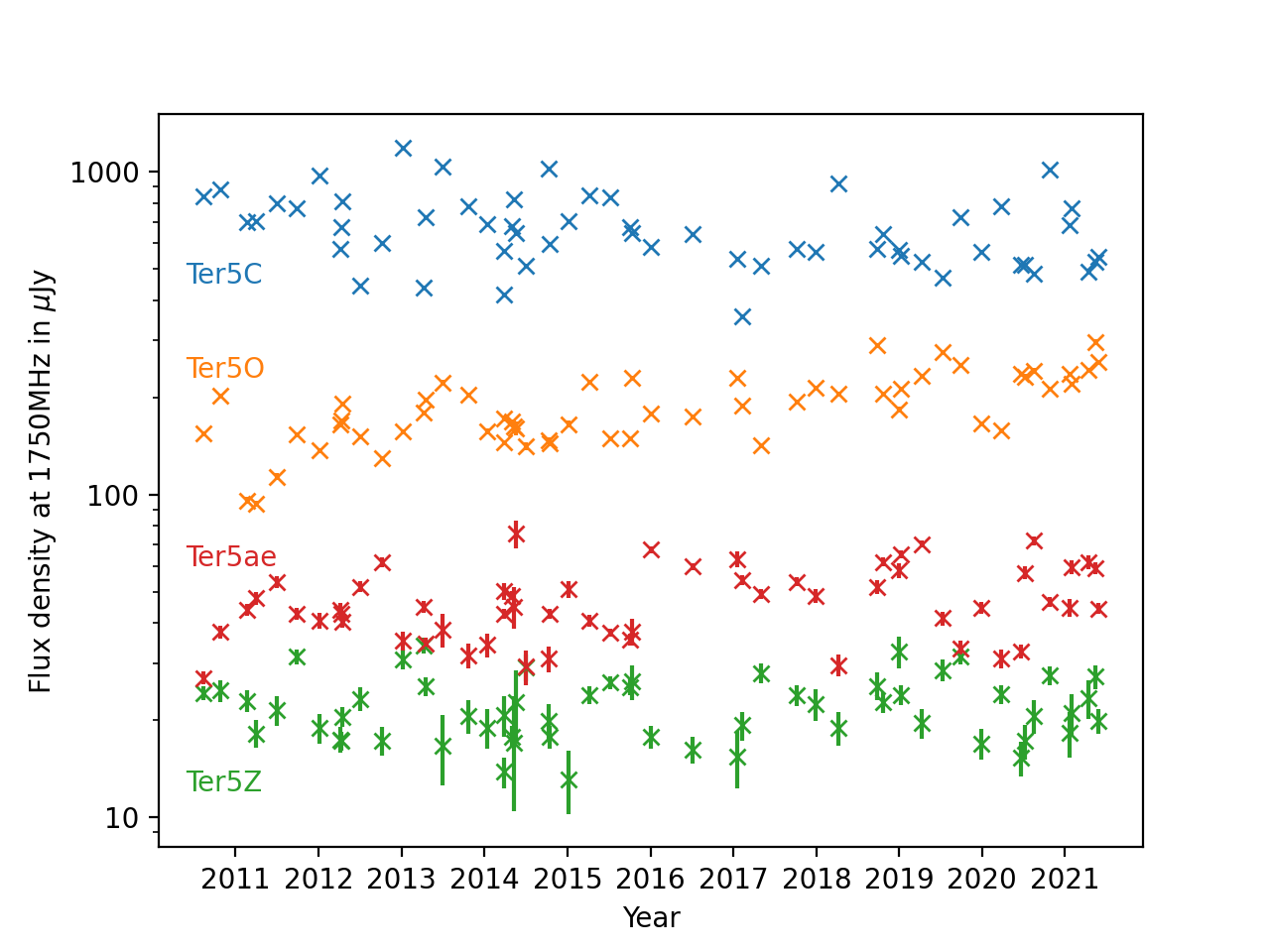}
        \caption{Calibrated flux densities at 1750\,MHz for Ter5C, O, ae, and Z, shown on a log scale (see \S\ref{ssec:SvT}). Statistical errors, when visible, are shown by a vertical bar. There is no obvious correlation between the pulsars, including for Ter5ae and Ter5Z, which are the closest pair of pulsars in \ter\ on the sky, with a 0.47$\arcsec$ separation.}
        \label{fig:Flux_density}
\end{figure}

Besides average flux densities, the large number of individual calibrated flux density measurements could inform refractive scintillation studies of Galactic field pulsars as well, since the fluxes vary on weeks to months timescales on the order of 10s to 100\% for each pulsar (see Fig.~\ref{fig:Flux_density}).  \citet{nt92} measured the scattering timescale $\tau_s$ at 685\,MHz to be $\sim$700\,$\mu$s, which corresponds to 16.4\,$\mu$s at 1750\,MHz assuming a scattering law $\propto\nu^{-4}$. The corresponding diffractive bandwidth \citep[e.g.~][]{rickett90,pulsar_handbook}, $\Delta f_\mathrm{DISS} \simeq 185\,\mathrm{Hz}\,(\tau_s / \mathrm{ms})^{-1} \sim 0.011$\,MHz, with a corresponding diffractive timescale $\delta T_\mathrm{DISS}$ of a few tens of seconds, meaning that we average over many thousands of scintles in each observation, ruling diffractive scintillation out as the main cause of the flux density variability. We are in the strong scintillation regime, though, with the scintillation strength $u = \sqrt{f/\Delta f_\mathrm{DISS}} \sim 400$, and the refractive timescale $\delta T_\mathrm{RISS} \sim u^2 \delta T_\mathrm{DISS}$ is weeks to months \citep{rickett96}. While month-long flux density correlations are not apparent in Fig.~\ref{fig:Flux_density}, observations separated by days to weeks do seem to be marginally correlated, at least by eye. The amount of modulation that we see, though, $m$=0.25$\pm$0.04, is fairly large given the predicted refractive modulation index, $m_\mathrm{RISS} \sim u^{-1/3} \sim 0.14$.  Future flux density studies of well-calibrated and monitored globular cluster pulsars seem warranted.

\section{Conclusion} \label{sec:conclusion}

We have created high S/N pulse profiles for 32 of the \ter\ pulsars, in both 1500\,MHz (L-band) and 2000\,MHz (S-band) bands. The profiles contain data taken by the Green Bank Telescope over 12 years which were summed together. In total the S-band profiles contain 124.9\,hours of observations and the L-band profiles contain 157.8\,hours. Due to instrumental issues, a large fraction of the GUPPI observations had corrupted polarization data, and were excluded from our polarization summation process. After excluding the corrupted data, the integrated polarization profiles had 71.3\,hours and 137.4\,hours in S-band and L-band, respectively.

From the summed polarization profiles, we determined rotation measures for 28 pulsars using \psrchive's {\tt{rmfit}} and 31 pulsars using {\tt{RMNest}}. The two methods in general agree with each other, and the pulsars that disagree between the two methods or are only measured via {\tt{RMNest}} have low amounts of linear polarization, are very faint pulsars, or have complicated pulse profile shapes. The RMs, along with DMs determined from the summed total intensity profiles, were combined to create a map of the projected and averaged parallel magnetic field strength over the spatial extent of \ter. We found that $\langle B_{||}\rangle$ has a rough gradient along the RA direction across the cluster of $\sim$6\,nG/arcsec ($\sim$160\,nG/parsec), or fractionally, a change of $\sim$20\%.

We measured the average flux densities for each of the 32 pulsars, along with their spectral indices. The flux densities ranged from one or two mJy for the brightest pulsar Ter5A, down to $\sim$10$\mu$Jy for the faintest known pulsars. Most of the other \ter\ pulsars have flux densities in the range of tens of $\mu$Jy. The average spectral index of the pulsars is approximately $-$1.35, which is flatter than the average of most pulsars known, likely due a selection effect. The pulsar luminosities inferred from these results appear to follow a power-law with a slope of $-$1 at the higher luminosities. Incompleteness at the lower luminosities suggests that \ter\ has dozens or even $>$100 additional pulsars waiting to be discovered.

Note: Polarization profiles of the pulsars in both bands are available from the authors upon request.

\begin{acknowledgments}
This project was supported by NSF award number 1852401, which funds the National Radio Astronomy Observatory / Green Bank Observatory Research Experience for Undergraduates. The Green Bank Observatory is a facility of the National Science Foundation operated under cooperative agreement by Associated Universities, Inc. The National Radio Astronomy Observatory is a facility of the National Science Foundation operated under cooperative agreement by Associated Universities, Inc. SMR is a CIFAR Fellow and is supported by the NSF Physics Frontiers Center awards 1430284 and 2020265. MED acknowledges support from the National Science Foundation (NSF) Physics Frontier Center award 1430284, and from the Naval Research Laboratory by NASA under contract S-15633Y. JWTH is supported in part by a Vici grant (VI.C.192.045) from NWO, the Dutch Research Council. Pulsar research at UBC is supported by an NSERC Discovery Grant and by the Canadian Institute for Advanced Research. We want to thank for the anonymous referee for comments which improved this paper.
\end{acknowledgments}

\vspace{5mm}
\facilities{GBT}

\software{{\tt ionFR} \citep{ionFR},
          {\tt PRESTO} \citep{presto}, 
          {\tt PSRCHIVE} \citep{psrchive_ascl},
          {\tt \tt RMNest} \citep{rmnest_ascl}}

\import{./}{table1.tex}

\bibliography{paper}{}
\bibliographystyle{aasjournal}

\end{document}

%% file: table1.tex
\begin{deluxetable*}{cccccccc}
\tablecaption{Terzan~5 Pulsar Properties\label{tab1}}
\tablehead{
    & \colhead{DM} & \colhead{RM$_{\tt rmfit}$} & \colhead{RM$_{\tt rmnest}$} &
    \colhead{$S_{\rm 1400}$} & \colhead{$S_{\rm 2000}$} & & \\
    \colhead{PSR}  & \colhead{(pc\,cm$^{-3}$)} & \colhead{(rad\,m$^{-2}$)} & 
    \colhead{(rad\,m$^{-2}$)} & \colhead{($\mu$Jy)} & \colhead{($\mu$Jy)} & \colhead{$\alpha$} & \colhead{$\chi_{\mathrm{red}}^2$} }
\startdata
 A & 242.34 & 191.2(3) & 193(2) & 2700 & 1700 & -1.31(15) & 310 \\
 C & 237.06 & 176.97(6) & 178.1(3) & 1100 & 670 & -1.4(2) & 337 \\
 D & 243.62 & 234(2) & 257(7) & 71 & 45 & -1.28(14) & 20 \\
 E & 236.63 & 171.0(7) & 171(4) & 170 & 110 & -1.16(13) & 23 \\
 F & 239.07 & 187.2(6) & 190(2) & 55 & 35 & -1.218(95) & 4.28 \\
 G & 237.34 & 196.4(4) & 195.79(98) & 24 & 22 & -0.26(11) & 10.4 \\
 H & 237.97 & 194(6) & 205(8) & 39 & 24 & -1.33(9) & 1.25 \\
 I & 238.55 & 181.7(7) & 183.1(1.5) & 95 & 55 & -1.53(11) & 22.9 \\
 J & 234.21 & 172.4(5) & 170.5(1.4) & 58 & 27 & -2.11(9) & 6.22 \\
 K & 234.46 & 175.0(1.0) & 175(4) & 66 & 39 & -1.47(7) & 0.577 \\
 L & 237.50 & 167(7) & 182(7) & 96 & 43 & -2.26(7) & 3.87 \\
 M & 238.49 & 169(2) & 179(4) & 140 & 91 & -1.14(11) & 22.1 \\
 N & 238.29 & 182.49(99) & 181(2) & 150 & 100 & -1.02(11) & 62.9 \\
 O & 236.20 & 176.2(3) & 174(2) & 310 & 160 & -1.9(2) & 38.9 \\
 Q & 234.24 & 175.4(8) & 179(5) & 56 & 36 & -1.24(14) & 4.64 \\
 R & 237.38 & 179(4) & 165.4(9.5) & 35 & 17 & -2.07(14) & 3.11 \\
 S & 236.22 & $-$ & 174(10) & 20 & 14 & -1.09(13) & 0.941 \\
 T & 237.80 & $-$ & $-$ & 26 & 15 & -1.61(9) & 0.66 \\
 U & 235.47 & $-$ & 238.6(9.5) & 30 & 12 & -2.491(97) & 1.69 \\
 V & 238.71 & 186.1(8) & 187(3) & 100 & 77 & -0.86(7) & 1.44 \\
 W & 238.92 & 174.3(9) & 183(7) & 54 & 31 & -1.53(14) & 12.6 \\
 X & 239.81 & 202(2) & 201(4) & 43 & 24 & -1.63(7) & 0.994 \\
 Y & 238.79 & 194(3) & 187(4) & 37 & 29 & -0.76(12) & 3.07 \\
 Z & 238.77 & $-$ & 201.3(9.8) & 30 & 23 & -0.7(2) & 2.1 \\
 aa & 237.43 & 206(7) & 158(22) & 29 & 20 & -1.00(14) & 2.2 \\
 ab & 238.40 & 171.2(1.1) & 171(3) & 45 & 23 & -1.83(12) & 0.754 \\
 ac & 238.69 & 203(3) & 204(7) & 31 & 17 & -1.67(15) & 1.28 \\
 ae & 238.61 & 184.3(1.1) & 175(5) & 56 & 50 & -0.3(2) & 7.88 \\
 af & 237.35 & 184(3) & 175(8) & 33 & 22 & -1.19(12) & 1.71 \\
 ag & 237.28 & 186(4) & 193(6) & 16 & 9.2 & -1.6(2) & 2.88 \\
 ah & 237.70 & 184(4) & 177(5) & 14 & 7.1 & -1.9(2) & 0.547 \\
 ai & 234.02 & 173.1(5) & 174.0(1.1) & 33 & 28 & -0.4(2) & 9.96 \\
\hline
\enddata

\tablecomments{Numbers in parentheses represent 1-$\sigma$ uncertainties
in the last digit(s). DM is the dispersion measure used to manually align
folded profiles as a function of frequency. RM$_{\tt rmfit}$ and RM$_{\tt
rmnest}$ are the rotation measures determined by {\tt PSRCHIVE}'s {\tt
rmfit} command and the {\tt rmnest} software, respectively. If there was
insufficient linear polarization for the software to measure the RM, a
``$-$'' is reported. $S_{1400}$ and $S_{2000}$ are the calibrated flux
densities at 1400\,MHz and 2000\,MHz, respectively, with statistical
errors of approximately 5\%, and systematic errors of $\sim$20\%. $\alpha$
is the fitted spectral index of the frequency ($\nu$) dependent flux
density, i.e.~$S_\nu \propto \nu^{\alpha}$, and the reduced-$\chi^2$ of
the fit is reported in the last column.}

\end{deluxetable*}